\documentclass{amsart}
\usepackage[centertags]{amsmath}
\usepackage{amsfonts}
\usepackage{amssymb}
\usepackage{amsthm}
\usepackage{newlfont}
\begin{document}

\title{Capacitors can radiate - some consequences of the two-capacitor problem with radiation}

\author{T.C. Choy}
\address{Centre for Materials Research University College London
Gower Street London WC1E 6BT}

\begin{abstract}
We fill a gap in the arguments of Boykin et al \cite{Boykin} by
not invoking an electric current loop ( i.e. magnetic dipole
model) to account for the radiation energy loss, since an obvious
corollary of their results is that the capacitors should radiate
directly even if the connecting wires are shrunk to zero length.
That this is so is shown here by a direct derivation of capacitor
radiation using an oscillating electric dipole radiator model for
the capacitors as well as the alternative less widely known
magnetic 'charge' current loop representation for an electric
dipole \cite{Schlekunoff}. Implications for Electromagnetic
Compliance (EMC) issues as well as novel antenna designs further
motivate the purpose of this paper.
\end{abstract}

\maketitle
\maketitle

\section{Introduction}
The recent paper by Boykin et al \cite{Boykin} on the
two-capacitor problem with radiation has a gap in their arguments
which we shall fill in this short note, which also serves to
provide an alternative and perhaps more direct derivation of their
key results. An important corollary of their results which we
shall show explicitly here implies that there can be radiation
{\it from the capacitors}. This radiation source is often not
discussed in many standard texts on Electromagnetism
\cite{Reitz,Panofsky,Griffiths} and in particular Antenna theory
\cite{Collin,Elliott,Kraus} and others, with the exception perhaps
of \cite{Schlekunoff} who highlighted the duality of magnetic
dipole radiation due to a physical current loop and the electric
dipole radiation due to a fictitious magnetic charge current loop
in his treatment. His seminal biconical antenna \cite{Schlekunoff}
is perhaps one of the few exactly solvable antenna models that
also demonstrates capacitor radiation. The importance of direct
radiation from capacitors has important implications for
Electromagnetic Compatibility (EMC) issues \cite {FCCHandbook} as
well as alternative non-conventional antenna designs currently
hotly debated in the engineering and amateur radio literatures
\cite{W5QJR,Hawker}. It is hoped that this note shall help to
clarify the basic physics involved in these alternative antennas.

We shall first discuss a corollary of the results of Boykin et al
\cite{Boykin}.  There is no doubt that their derivation
demonstrates that the missing energy can be accounted for by
radiation as they have shown using a magnetic dipole, current loop
model.  We can then ask the question as to what would happen if
the current loop is shrunk to an infinitely small radius.  Their
result then indicates that the capacitors must now become the
source of radiation.  This can be seen in two ways.  1) In their
derivation eqn(16) indicates that the physical dimension of the
current loop b (which is absorbed in the constant K) eventually
cancels out so that in fact the limit $b \rightarrow 0$ can be
taken in eqn(16) without mathematical difficulties. 2) Owing to an
equivalence theorem between a circular and a square loop radiator
(see their Fig 1) of identical area \cite{Kraus}, we can now
shrink the wire dipole radiators to zero length, from which energy
conservation would imply that the capacitors must themselves serve
as radiators. That this is so will be demonstrated in the next two
sections.
\section{Oscillating electric dipole model}.
In the long wavelength limit, the two capacitors connected in
parallel with zero length wires can be viewed as an oscillating
electric dipole due to a series capacitor of value $C_s$, (see Fig
3 of \cite{Boykin}) . The power intensity of radiation from such a
dipole of moment p is now given by \cite{Landau}:
\begin{equation}
P_{rad} = \frac{1}{6\pi\epsilon_0 c^3}\ddot{p}^2.
\end{equation}
In our ideal capacitor model considered as two parallel plates
with separation $\ell$, then:
\begin{equation}
\ddot{p} = C_s \ddot{V_c}^2 \ell^2 .
\end{equation}
Since now:
\begin{equation}
V_X =
\frac{P_{rad}}{I}=\frac{P_{rad}}{\dot{Q}}=\frac{P_{rad}}{\dot{V}_c
C_s},
\end{equation}
the non-linear differential equation eqn(12) of \cite{Boykin}
immediately follows in a similar way using their lump circuit
model to account for the radiation resistance $X$ i.e.:
\begin{equation}
\ddot{V}_c^2 + \frac{1}{K_C C_s} \dot{V_c} V_c = 0 . \label{NNDE}
\end{equation}
except that now we have:
\begin{equation}
K_C=\frac{\ell^2}{6\pi\epsilon_0 c^3}.
\end{equation}
The rest of the proof for the radiation energy follows identically
as \cite{Boykin} so that we do not need to reproduce them here.
This result can also be derived using an analogous model as
\cite{Boykin}, but this time using a fictitious magnetic current
loop model \cite{Schlekunoff}.  As these formulas are not often
used in standard texts we shall provide the details in the next
section, highlighting the advantage that this model will be more
useful in terms of evaluating actual antenna radiation
characteristics in more realistic capacitor antennas using
standard formulas. Before we do so it is perhaps worth drawing to
attention that the point dipole model is an extreme limit for the
capacitor since like the corresponding short current dipole for
wires \cite{Reitz,Panofsky,Griffiths,Collin,Elliott,Kraus}, and as
for long wires \cite{Kraus} of order $\lambda$, the contributions
should be added vectorially for each element and integrated over
the capacitance area for a $\lambda$ size capacitor.
\section{Magnetic current loop model}
The magnetic current loop model uses the fact that Faraday's law
of magnetic induction can be used to define a magnetic 'charge'
current as the source for an electric dipole field.  In this case
we shall have a not frequently used vector potential {\bf F} such
that:
\begin{equation}
curl\ {\mathbf F} = - {\mathbf D}, \quad {\mathbf
H}=-\frac{\partial {\mathbf F}}{\partial t};
\end{equation}
and thus also:
\begin{equation}
{\mathbf F} = \frac{\epsilon_0}{4\pi}\int \frac{I_M(t-r/c)}{r} dl,
\end{equation}
where $I_M$ specifies the magnetic current of the loop
\cite{Funits}. Once again the results of \cite{Boykin} applies by
analogy in particular their eqn(8), upon replacing $\epsilon_0$ by
$\mu_0$:
\begin{equation}
P_{rad} = \frac{\pi b_m^4}{6\mu_0 c^5} [\ddot{I}_M(t-r/c)]^2,
\end{equation}
where $b_m$ is now the magnetic current loop radius which should
be at least the radius of the assumed circular parallel plate
capacitors. In view of Faraday's law $V_c=-I_M$:
\begin{equation}
V_X = \frac{P_{rad}}{I}=K_M \frac{\ddot{V}_c^2}{C_s \dot{V}_c},
\end{equation}
and once again eqn(\ref{NNDE}) follows except that now $K_M$ is
defined as:
\begin{equation}
K_M=\frac{\pi b_m^4}{6\mu_0 c^5},
\end{equation}
which as we can see is a less efficient radiator.  For the same
size loop $b_m=b$ the ratio:
\begin{equation}
\frac{K}{K_M} \approx 10^5.
\end{equation}
The implications of this result for EMC is also important. Since
in terms of spectral content, a less efficient radiator would in
the capacitor system tend to spread the energy over a wider
spectrum, i.e. assuming ideal capacitors with no internal losses.
Both magnetic or electric dipoles have an intensity that goes as a
fourth power of frequency $\omega^4$ \cite{Landau}. Indeed as
noted in \cite{Boykin} the radiation resistance is given by
$R_{rad}=K s^4$ in the frequency domain. Hence for a smaller $K$
the radiation time constant $\tau=R_{rad}C_s$ is smaller which
implies a wider spread in radiation energy. Note however that the
point dipole model of the previous section yields a different
picture. Here the efficiency factor is given by:
\begin{equation}
\frac{K}{K_C}=\frac{2\pi^2}{c^2}\frac{b^4}{\ell^2}\approx 10^{-15}
\frac{b^4}{\ell^2}>\approx 10^7 C_s^2.
\end{equation}
which for large capacitances can be comparable to the wire loop.
Practical capacitance antennas, depending on frequencies will
behave somewhere between a point electric dipole versus a magnetic
current loop, as we shall see in the next section \cite{wireloop}.
\section{Capacitance antennas}
In recent years there has been controversies in the engineering
community regarding certain capacitance antennas, patented in the
US and Britain \cite{W5QJR,Hawker}, which purports to use Poynting
vector synthesis for its operational principles.  These are in
fact now commercial products that have received contradicting
support from broadcasting applications. We do not however
subscribe to the theory of Poynting vector synthesis
\cite{W5QJR,Hawker}. Nevertheless, the analysis from the last
section shows that (a) capacitance radiation is a reality and (b)
their efficiencies based on the idealized magnetic 'charge'
current loop model which is close to their practical counterparts,
depend very much on the capacitance disc sizes. To achieve similar
efficiency as a wire loop, the capacitance antenna loop $b_m$
needs to be an order of magnitude bigger than the magnetic loop
area b at low frequencies:
\begin{equation}
b_m=10 b.
\end{equation}
It would seem that any low profile advantages achieved in the use
of capacitance antennas has to be paid for by much larger
capacitance disc areas.  However at high frequencies in which the
point electric dipole model is approached, capacitors might be
better radiators, which might in fact find useful applications for
the new digital spread spectrum modes of transmission. Having said
this, most antennas are not meant to act as broadband radiators.
Wire antennas resonate at the operating frequency by making use of
either free space capacitances leading to the empirical formula
\cite{Collin,Elliott,Kraus}:
\begin{equation}
L=\frac{143}{f_{Mhz}},
\end{equation}
for the length of a half-wave dipole antenna in metres up to
several hundred MHz. In the same way capacitance antennas can
resonate using free space inductances. The corresponding formula
for the capacitance dipole, including its relative performance
would be an interesting exercise for the student. For practical
systems a lump circuit analysis \cite{Boykin} by introducing an
inductance L which could be both a sum of stray plus external
inductances would suffice. However modern Pspice software and
various antenna modelling software do not include the radiation
resistance from capacitances discussed here, so some care needs to
be exercised in their use.
\section{Conclusion}
We have filled a gap in the discussion of the radiation from the
transient switching of charges between two capacitors.  We showed
that without a wire loop, or in the limit where the wire lengths
are infinitesimal, the capacitance system will radiate, using a
point electric dipole model or more appropriately a magnetic
'charge' current loop model.  These results should be added to
modern texts on electromagnetism and antenna theory. The
implications of our results for EMC directives and for novel
antenna designs are significant and should also be noted in
physics and engineering teaching courses.
\bibliographystyle{amsplain}
\bibliography{TwoCapacitor.bbl}

\begin{thebibliography}{10}
\bibitem{Funits} The SI units for magnetic 'charge' current is in fact in volts.
\bibitem{wireloop} This property is of course similar to wire antennas.
\end{thebibliography}

\end{document}